\documentclass[letterpaper, 10 pt, conference]{ieeeconf}  % Comment this line out
\IEEEoverridecommandlockouts                              % This command is only
\usepackage[utf8]{inputenc}
\usepackage[T1]{fontenc}
\usepackage{graphicx}
\usepackage{comment}
\usepackage{placeins}
\usepackage[hidelinks]{hyperref}

\usepackage{array}
\makeatletter
\newcommand{\thickhline}{%
    \noalign {\ifnum 0=`}\fi \hrule height 1pt
    \futurelet \reserved@a \@xhline
}
\newcolumntype{"}{@{\hskip\tabcolsep\vrule width 1.5pt\hskip\tabcolsep}}
\makeatother

\usepackage{tikz,xcolor,hyperref}

\definecolor{lime}{HTML}{A6CE39}
\DeclareRobustCommand{\orcidicon}{%
	\begin{tikzpicture}
	\draw[lime, fill=lime] (0,0) 
	circle [radius=0.16] 
	node[white] {{\fontfamily{qag}\selectfont \tiny ID}};
	\draw[white, fill=white] (-0.0625,0.095) 
	circle [radius=0.007];
	\end{tikzpicture}
	\hspace{-2mm}
}

\foreach \x in {A, ..., Z}{%
	\expandafter\xdef\csname orcid\x\endcsname{\noexpand\href{https://orcid.org/\csname orcidauthor\x\endcsname}{\noexpand\orcidicon}}
}

\title{\LARGE \bf Classification of Electrical Impedance Tomography Data Using Machine Learning}

\author{Diogo Pessoa$^{1}$\orcidA{}, Bruno Machado Rocha$^{1}$\orcidB{}, Grigorios-Aris Cheimariotis$^{2}$\orcidC{}, Kostas Haris$^{2}$\orcidD{},\\ Claas Strodthoff$^{3}$\orcidE{}, Evangelos Kaimakamis$^{2,4}$\orcidF{}, Nicos Maglaveras$^{2}$\orcidG{}, Inéz Frerichs$^{3}$\orcidH{},\\ Paulo de Carvalho$^{1}$\orcidI{} and Rui Pedro Paiva$^{1}$\orcidJ{}% <-this % stops a space

\thanks{$^{1}$University of Coimbra, Centre for Informatics and Systems of the University of Coimbra, Department of Informatics Engineering, 3030-290 Coimbra, Portugal. {\tt\small \{dpessoa, bmrocha, carvalho, ruipedro\}@dei.uc.pt}}
\thanks{$^{2}$ Laboratory of Computing, Medical Informatics and Biomedical Imaging Technologies, Aristotle University of Thessaloniki, Thessaloniki, 54636, Greece. {\tt\small \{ncheimar, kharis, nicmag\}@auth.gr}}
\thanks{$^{3}$ Department of Anesthesiology and Intensive Care Medicine, University Medical Centre Schleswig-Holstein, Campus Kiel, Kiel,
Germany. {\tt\small \{Claas.Strodthoff, Inez.Frerichs\}@uksh.de}}
\thanks{$^{4}$ 1st Intensive Care Unit, “G. Papanikolaou” General Hospital, Thessaloniki, Greece. {\tt\small vkaimak@med.auth.gr}}
}%

\begin{document}

\maketitle
\thispagestyle{empty}
\pagestyle{empty}
%%%%%%%%%%%%%%%%%%%%%%%%%%%%%%%%%%%%%%%%%%%%%%%%%%%%%%%%%%%%%%%%%%%%%%%%%%%%%%%%
\begin{abstract}
Patients suffering from pulmonary diseases typically exhibit pathological lung ventilation in terms of homogeneity. Electrical Impedance Tomography (EIT) is a non-invasive imaging method that allows to analyze and quantify the distribution of ventilation in the lungs. In this article, we present a new approach to promote the use of EIT data and the implementation of new clinical applications for differential diagnosis, with the development of several machine learning models to discriminate between EIT data from healthy and non-healthy subjects. EIT data from 16 subjects were acquired: 5 healthy and 11 non-healthy subjects (with multiple pulmonary conditions). Preliminary results have shown accuracy percentages of 66\% in challenging evaluation scenarios. The results suggest that the pairing of EIT feature engineering methods with machine learning methods could be further explored and applied in the diagnostic and monitoring of patients suffering from lung diseases. Also, we introduce the use of a new feature in the context of EIT data analysis (Impedance Curve Correlation).
\end{abstract}

\section{Introduction}
Lung-related diseases represent some of the most common medical conditions worldwide \cite{Rocha2021}. Moreover, they are among the most significant causes of morbidity and mortality and are responsible for a substantial strain on health systems \cite{WHO2018,Gibson2013}. Most of these diseases can be characterized by increased ventilation inhomogeneity associated with the pathologically changed regional lung function. Therefore, parameters that translate these changes in lung function can be useful in the diagnostic and monitoring of such diseases. 

Electrical Impedance Tomography (EIT) is a non-invasive, radiation-free, imaging technique that relies on the application of alternating electrical currents on the external surface of the body to assess its internal electrical characteristics and generate bio-impedance images/maps \cite{Khan2019,Krauss2021}. Hence, the reconstructed images represent the electrical permittivity and conductivity distributions inside the chest \cite{Khan2019}.

Despite the lower spatial resolution of EIT when compared to other traditional well established imaging modalities, such as magnetic resonance imaging and computed tomography scan, EIT can provide continuous monitoring and it has a high temporal resolution, unlike the other methods \cite{Khan2019}. Therefore, while traditional imaging techniques can provide further insight regarding structural changes of the pulmonary tissue, EIT is better suited to assess its functionality \cite{Lasarow2021}.  

Over the years, the number of studies focusing on EIT has experienced a steady increase. However, the majority of these have not aimed at the development of methods to promote differential diagnosis applications with EIT. This could represent one of the greatest strengths of EIT, especially due to its suitability to be deployed in wearable devices, allowing to closely monitor patients with telemonitoring systems \cite{Frerichs2020}. 

In \cite{Trenk2016}, the authors suggested, through statistical inference, that the global inhomogeneity index extracted from EIT might be useful for the identification and follow up of ventilation problems in patients with COPD. Also, in \cite{Krauss2021}, the authors have assessed, through the extraction and analysis of several measures, the usefulness of EIT in the characterization of regional ventilation in idiopathic pulmonary fibrosis. In another study, the authors have used raw EIT images paired with regression machine learning techniques to infer several respiratory and circulatory parameters \cite{Strodthoff2021}. 
%In another study, the authors have also used this index to evaluate its utility to predict failure of a spontaneous breathing trial \cite{Bickenbach2017}.

The main objective of this exploratory study is to understand if the association of EIT and machine learning could be useful for differential diagnosis applications, through the employment of multiple traditional machine learning classification models. To the best of our knowledge, this is the first study where EIT data is employed in association with a machine learning classification pipeline, for this purpose. Therefore, we studied whether EIT derived measures of spatial and temporal ventilation homogeneity, extracted from periods of tidal and deep breathing, were able to discriminate between EIT data coming from healthy and non-healthy subjects. 

%----> Present motivation and main objectives and results of the work
%This data allowed us to examine whether EIT measures of ventilation heterogeneity derived from periods of tidal and deep breathing were able do discriminate among healthy and non-healthy subjects.
\section{Materials and Methods}
\subsection{Data acquisition and protocol}
The data collection process for this study was carried out in the pneumology service at the George Papanikolaou General Hospital of Thessaloniki, Greece. The study was approved by the Ethics Committee of the same hospital.

A total of 16 (6 female, 10 male) participants were considered in this study, which included  5 healthy and 11 non-healthy subjects. The group of non-healthy subjects suffered from a wide range of conditions (such as COPD, asthma, pneumonia, etc.). The average age for all subjects was 60.3$\pm$20.4 years with a mean BMI of 28.9$\pm$5.5 $kg/m^{2}$ (Healthy - 35.3$\pm$14.4 years and 31.0$\pm$6.5 $kg/m^{2}$; Non-healthy - 71.6$\pm$9.6 years and 28.0$\pm$5.1 $kg/m^{2}$).

EIT data were collected using the Goe-MF II EIT device (CareFusion, Höchberg, Germany). An array of sixteen self-adhesive electrodes (Blue Sensor, Ambu, Ballerup, Denmark) was attached to the chest circumference between the 5-6th intercostal space (xipho-sternal line), with another reference electrode placed on the abdomen. Small alternating electrical currents were delivered through adjacent pairs of electrodes in a sequential rotating process and the resulting potential differences were measured by the remaining electrode pairs. EIT data were acquired at a sampling rate of 33 images/second.

%During the measurement, small alternating electrical currents were delivered and measured by all electrodes alternately (50 kHz, 5 mA peak-to-peak). The currents were applied through adjacent pairs of electrodes in a sequential rotating process and the resulting potential differences were measured by the remaining electrode pairs. EIT data were acquired at a sampling rate of 33 images/second.

During the acquisitions the participants were asked to perform two types of recordings: 1) tidal breathing during two or three respiratory cycles, followed by deep breathing, also during two or three respiratory cycles; 2) tidal breathing followed by a period of about two minutes with both harms elevated above the head. Each participant has repeated the two types of recordings multiple times.

\subsection{EIT ROIs}\label{subsec:EITrois}
\begin{figure}[ht!]
	\centering
	\includegraphics[trim={0.6cm 0.1cm 0.6cm 0.2cm},clip,width=\columnwidth]{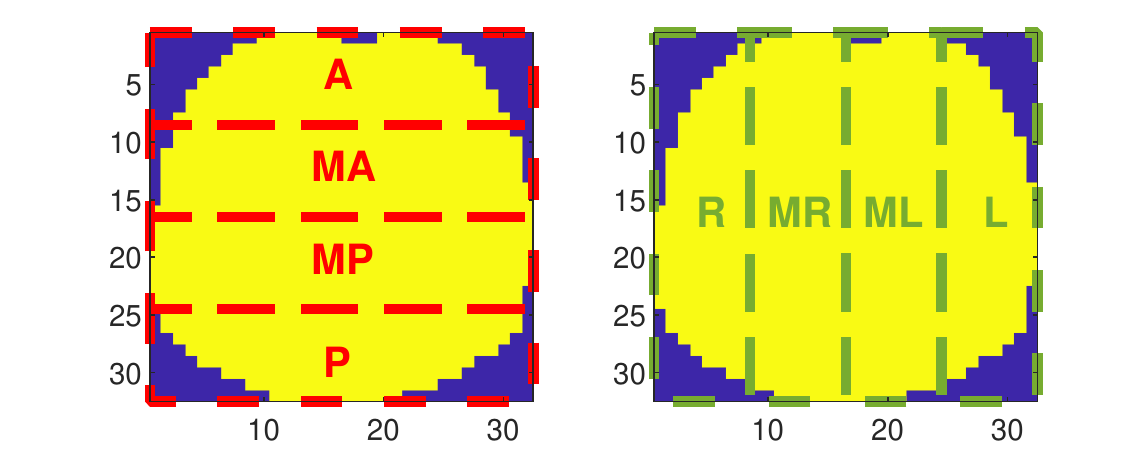}
	\caption{EIT horizontal and vertical ROIs (Left - horizontal ROIs, Right - vertical ROIs; A-Anterior; P-Posterior; R-Right; L-Left; M-Mid).}
	\label{fig:EITrois}
\end{figure}

In order to take advantage of the EIT capability to perform regional ventilation analysis \cite{Frerichs2016}, several major Regions of Interest (ROIs) were considered to extract meaningful parameters from the data. Therefore, based on the model used to obtain the reconstructed images (\autoref{subsec:EITreconstruction}), the images were divided in multiple regions. A total of 16 different regions was considered. These include major ROIs, such as the right/left and anterior/posterior areas of the lung, as well as 4 different quadrants and vertical and horizontal ROIs. In \autoref{fig:EITrois}, the vertical and horizontal ROIs are represented, as well as the shape of the model used by the EIDORS library (yellow area). 
%The upper section of the images represents the anterior area of the lung and the lower part the posterior area, accordingly. Lastly, the left region of the image represents the right lung, and vice-versa.

\subsection{EIT reconstruction}\label{subsec:EITreconstruction}
\begin{figure}[ht!]
	\centering
	\includegraphics[trim={0.2cm 0cm 0.5cm 0.2cm},clip,width=0.95\columnwidth]{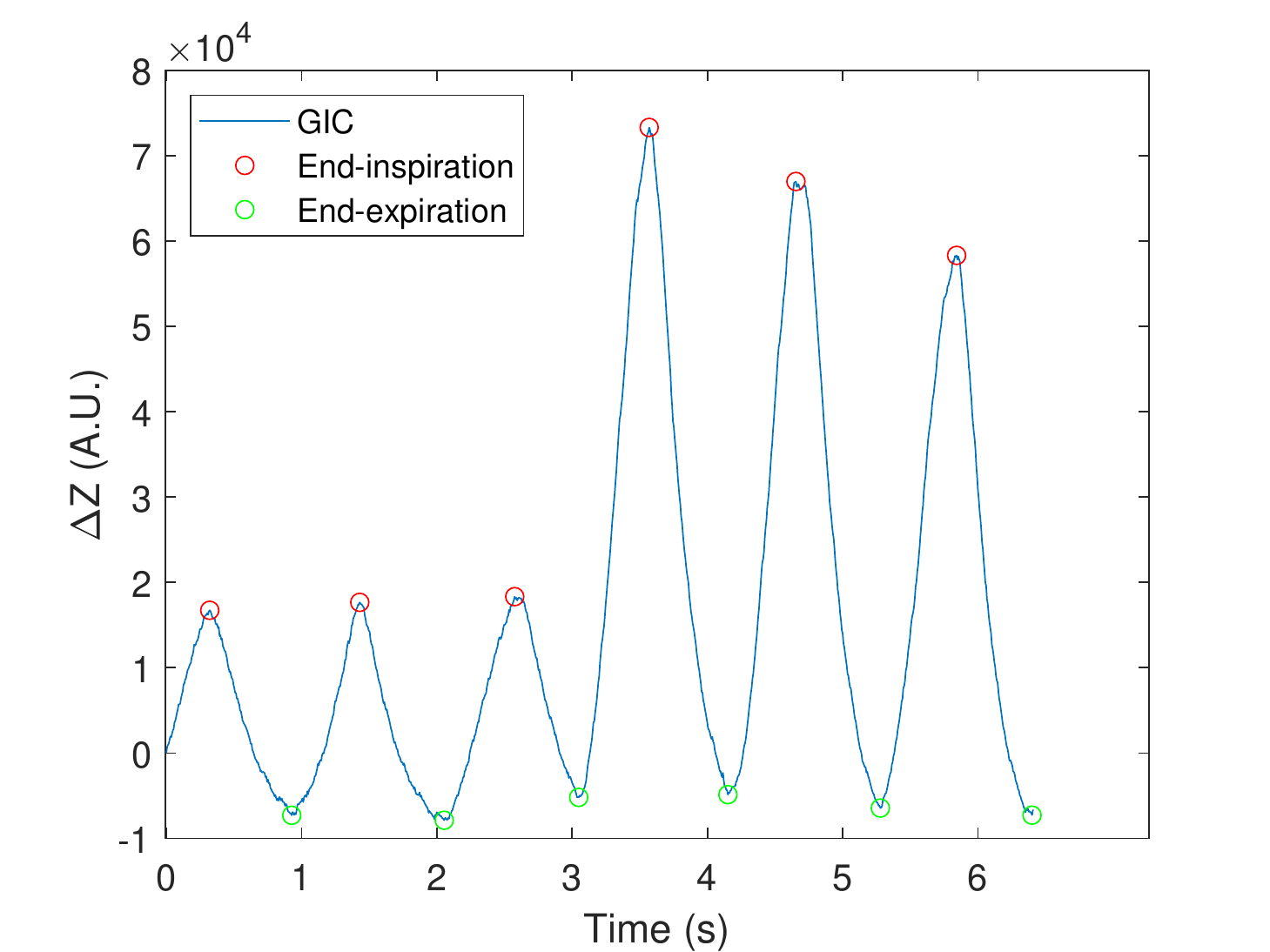}
	\caption{Example of GIC with the identification of end-inspiratory and end-expiratory moments (A.U.-Arbitrary Units; $\Delta Z$-Impedance variation).}
	\label{fig:GICexample}
\end{figure}

The acquired raw EIT data were processed offline to obtain the reconstructed images/frames using the Graz Consensus Reconstruction Algorithm for EIT (GREIT) \cite{Adler2009}. The reconstruction was performed using an adult thorax shaped model with a single plane of 16 electrodes and the adjacent stimulation pattern was selected from the models library of the EIDORS software \cite{Adler2006}. The resulting reconstructed EIT images consisted of 32 by 32 pixels. After obtaining the reconstructed images for every time step, the Global Impedance Curve (GIC) was computed through the sum of all individual pixel values for each image. Using the GIC, the end-inspiration and end-expiration moments were manually determined/annotated through the analysis of the curves of every EIT file considered (see \autoref{fig:GICexample}).
%\footnote{http://eidors3d.sourceforge.net/}

%\FloatBarrier

\begin{figure}[ht!]
  \centering
  \includegraphics[width=0.9\columnwidth]{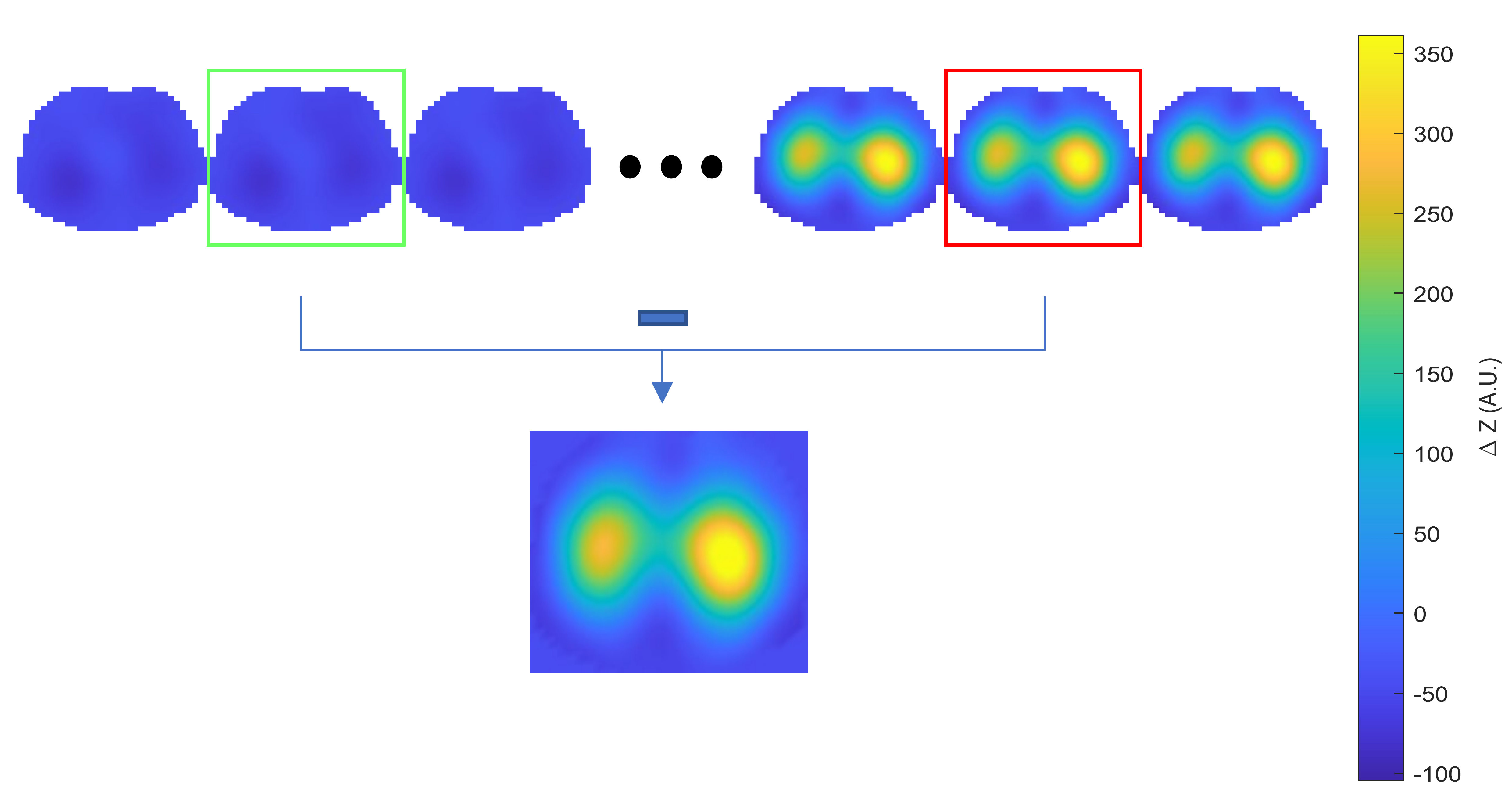}
  \caption{Functional EIT image generation: difference between end-inspiration (red square) and begin-inspiration (green square) frames (A.U.-Arbitrary Units; $\Delta Z$-Impedance variation).}
  \label{fig:fEIT}
\end{figure}

After the determination of the respiratory cycles, the functional EIT images were computed (fEIT images). These images were determined for each breathing cycle and correspond to the difference of the EIT reconstructed images between end-inspiration and end-expiration instants (see \autoref{fig:fEIT}). The regional impedance curves (RIC) for each of the considered ROIs (see \autoref{subsec:EITrois}) were also isolated between the above-mentioned moments.

\subsection{EIT measures/features}\label{subsec:EITmeasures}
In this study, several EIT features were extracted at a global and regional level, based on the regions defined in \autoref{subsec:EITrois}. The features were extracted for each inspiration cycle individually, from each fEIT image and the corresponding impedance curves (as identified in \autoref{subsec:EITreconstruction}). A total of 112 features were extracted for each sample using methods developed in MATLAB 2020a. A brief description of the extracted parameters/features is presented below:
\begin{itemize}
    
    \item \textit{Ventilation Ratios:} ratio between the sum of all pixel values from two certain regions (e.g., ratio between left and right lungs, etc) \cite{Frerichs2016};
    
    \item \textit{Coefficient of Variation:} statistical measure that characterizes the relative magnitude of the standard deviation of a fEIT image with respect to its mean \cite{Frerichs2016}. This measure was determined for the multiple ROIs;
    
    \item \textit{Global Inhomogeneity Index:} measure that quantifies the homogeneity of the volume distribution \cite{Zhao2014}. The higher it its, the more heterogeneous the ventilation distribution is. This measure was determined for the multiple ROIs;
    
    \item \textit{Regional Ventilation Delay:} the regional ventilation delay expresses the delay between the global start of inspiration and the point in time where the regional impedance curve reaches a certain threshold (set to 40\%) \cite{Wrigge2008}. Thus, it can be used to translate the temporal heterogeneity of ventilation. This measure is determined for multiple regional impedance curves in relation to the GIC;
    
    %\item Intertidal Gas Distribution:
    
    \item \textit{Impedance Curve Correlation:} The impedance curve correlation is determined through the correlation between the GIC and the regional impedance curves, and between the regional impedance curves within themselves. This metric captures the dependency in variation of the impedance curves, translating the spatial and temporal homogeneity of ventilation in lung function.
\end{itemize}

\subsection{Classifiers}
To develop the classification models, we used several machine learning algorithms to classify the samples, such as: Linear discriminant analysis (LDA); Support vector machine with radial basis function kernel (SVMrbf); Decision Trees (DecTree); Random undersampling boosted trees (RUSBoost); and Random Forest (RndForest). All the classifiers were trained 50 times with different seeds to divide the complete dataset, and their hyperparameters were optimized on a validation set containing 25\% of the training set. Bayesian optimization was used to optimize the models' hyperparameters \cite{Snoek2012}. The models with the best set of hyperparameters were then applied to the test set. Filter feature selection algorithms (e.g., ReliefF and MRMR) were tested but did not lead to a better performance of the algorithms. Therefore, the results for these cases are not presented. All the models were developed in MATLAB 2020a.

%\begin{comment}
%\subsection{Feature Selection}
%After preliminary experiments, the minimum redundancy maximum relevance (MRMR) algorithm was chosen to perform feature selection. The MRMR algorithm finds an optimal set of features that is mutually and maximally dissimilar and can represent the response variable effectively. The algorithm minimizes the redundancy of a feature set and maximizes its relevance to the response variable \cite{Ding2005}. 
%\end{comment}

\section{Results and Discussion}

In this section, we present the performance evaluation of the algorithms. All the classification methods were tested in two different scenarios: A) Repeated stratified hold-out validation; B) Repeated stratified patient hold-out validation. 

Scenario A consists in the random division of all the samples available with 75\% of the data being used to train the models and the remaining 25\% for testing. This is a scenario frequently found in many research approaches, which however has significant limitations as will be explained later in this section. On the other hand, in scenario B the division occurs patient wise, with 75\% of the total number of patients used in the training set and the remaining 25\% in the testing set, accordingly. The patient-wise split guarantees that data from the same patient cannot be in the training and testing datasets at the same time. This split is performed in such a way that the number of available samples on the testing set is always equal, or superior, to 25\% of the total number of available samples. Both processes are repeated 50 times, being that 25\% percent of the training data is retained and used for validation to perform hyperparameter optimization during the training phase, as previously mentioned.

To develop the machine learning classification models a total of 1500 samples were considered (1108 from non-healthy subjects and 392 from healthy ones, respectively). In \autoref{tab:resultsHoldOutVal} and \autoref{tab:resultsPatientHoldOutVal} the results obtained over 50 repetitions for both evaluation scenarios are presented.

\begin{table*}[ht!]
\centering
\caption{Classification performance evaluation for Repeated hold-out validation (Scenario A).}
\label{tab:resultsHoldOutVal}
\begin{tabular}{|c"c"c|c|c|c"c|c|c|c|}
\hline
\textbf{Class} & \textbf{Acc} & \textbf{F1 H} & \textbf{Rec H} & \textbf{Prec H} & \textbf{Spec H} & \textbf{F1 NH} & \textbf{Rec NH} & \textbf{Prec NH} & \textbf{Spec NH} \\ \hline

DecisionTree & 90.9$\pm$1.4 & 80.8$\pm$4.1 & 84.0$\pm$4.3 & 94.5$\pm$1.7 & 82.3$\pm$2.8 & 94.5$\pm$1.7 & 93.3$\pm$1.3 & 80.8$\pm$4.1 & 93.9$\pm$1.0 \\ \hline
RUSBoost & 95.1$\pm$1.5 & 89.0$\pm$3.9 & 92.1$\pm$4.2 & 97.2$\pm$1.7 & 90.4$\pm$2.9 & 97.2$\pm$1.7 & 96.2$\pm$1.3 & 89.0$\pm$3.9 & 96.7$\pm$1.1 \\ \hline
LDA & 86.2$\pm$4.1 & 77.6$\pm$5.1 & 73.8$\pm$9.6 & 89.3$\pm$6.6 & 75.0$\pm$4.8 & 89.3$\pm$6.6 & 91.9$\pm$1.5 & 77.6$\pm$5.1 & 90.4$\pm$3.4 \\ \hline
%SVMlin & 91.2$\pm$1.9 & 85.1$\pm$5.2 & 82.2$\pm$4.3 & 93.4$\pm$2.0 & 83.5$\pm$3.7 & 93.4$\pm$2.0 & 94.7$\pm$1.8 & 85.1$\pm$5.2 & 94.0$\pm$1.3 \\ \hline
SVMrbf & 95.7$\pm$1.2 & \textbf{89.2$\pm$4.0} & 94.4$\pm$2.9 & 98.1$\pm$1.1 & 91.6$\pm$2.5 & 98.1$\pm$1.1 & \textbf{96.3$\pm$1.3} & \textbf{89.2$\pm$4.0} & 97.2$\pm$0.8 \\ \hline
RndForest & \textbf{96.0$\pm$1.2} & 88.7$\pm$3.8 & \textbf{96.0$\pm$3.7} & \textbf{98.6$\pm$1.4} & \textbf{92.1$\pm$2.4} & \textbf{98.6$\pm$1.4} & \textbf96.1$\pm$1.2 & 88.7$\pm$3.8 & \textbf{97.3$\pm$0.8} \\ \hline

\end{tabular}
\begin{tabular}{@{}c@{}} 
\multicolumn{1}{p{0.9\textwidth}}{\footnotesize \textit{Acc}-Accuracy; \textit{F1}-F1 Score; \textit{Rec}-Recall; \textit{Prec}-Precision; \textit{Spec}-Specificity; \textit{H}-Healthy; \textit{NH}-Non-healthy.}
\end{tabular}
\end{table*}

\begin{table*}[ht!]
\centering
\caption{Classification performance evaluation for repeated patient hold-out validation (Scenario B).}
\label{tab:resultsPatientHoldOutVal}
\begin{tabular}{|c"c"c|c|c|c"c|c|c|c|}
\hline
\textbf{Class} & \textbf{Acc} & \textbf{F1 H} & \textbf{Rec H} & \textbf{Prec H} & \textbf{Spec H} & \textbf{F1 NH} & \textbf{Rec NH} & \textbf{Prec NH} & \textbf{Spec NH} \\ \hline

DecisionTree & 62.1$\pm$8.7 & 26.9$\pm$13.0 & 46.4$\pm$20.6 & 81.0$\pm$11.6 & 32.4$\pm$13.0 & 81.0$\pm$11.6 & 67.4$\pm$6.7 & 26.9$\pm$13.0 & 73.3$\pm$7.5 \\ \hline
RUSBoost & 65.1$\pm$8.1 & \textbf{35.5$\pm$15.9} & 50.9$\pm$19.8 & 80.8$\pm$10.8 & \textbf{40.3$\pm$15.4} & 80.8$\pm$10.8 & \textbf{70.3$\pm$6.3} & \textbf{35.5$\pm$15.9} & 74.8$\pm$6.7 \\ \hline
LDA & 62.3$\pm$8.4 & 25.7$\pm$16.0 & 52.6$\pm$26.3 & 82.2$\pm$14.8 & 30.3$\pm$14.6 & 82.2$\pm$14.8 & 67.5$\pm$6.0 & 25.7$\pm$16.0 & 73.4$\pm$7.5 \\ \hline
%SVMlin & 57.8$\pm$7.3 & 30.6$\pm$14.2 & 37.7$\pm$15.8 & 72.5$\pm$10.7 & 32.5$\pm$13.1 & 72.5$\pm$10.7 & 66.3$\pm$5.9 & 30.6$\pm$14.2 & 68.8$\pm$6.4 \\ \hline
SVMrbf & 59.8$\pm$8.5 & 22.6$\pm$10.4 & 41.6$\pm$19.6 & 79.8$\pm$12.4 & 27.7$\pm$10.7 & 79.8$\pm$12.4 & 65.7$\pm$6.1 & 22.6$\pm$10.4 & 71.7$\pm$7.6 \\ \hline
RndForest & \textbf{66.0$\pm$7.3} & 28.7$\pm$15.3 & \textbf{54.2$\pm$22.8} & \textbf{85.9$\pm$11.0} & 35.4$\pm$15.9 & \textbf{85.9$\pm$11.0} & 69.4$\pm$5.5 & 28.7$\pm$15.3 & \textbf{76.4$\pm$6.2} \\ \hline

\end{tabular}
\begin{tabular}{@{}c@{}} 
\multicolumn{1}{p{0.9\textwidth}}{\footnotesize \textit{Acc}-Accuracy; \textit{F1}-F1 Score; \textit{Rec}-Recall; \textit{Prec}-Precision; \textit{Spec}-Specificity; \textit{H}-Healthy; \textit{NH}-Non-healthy.}
\end{tabular}
\end{table*}

Through the analysis of both tables, the disparity between the two scenarios is evident. In scenario A, all classifiers performed well, with the random forest model presenting the highest score, reaching an average accuracy of 96.0$\pm$1.2\%. On the other hand, in scenario B, the performance drops significantly for all models and, once again, the random forest model presents the highest score, reaching an average accuracy of 66.0$\pm$7.3\%. From the analysis of \autoref{tab:resultsPatientHoldOutVal}, it is possible to observe that the performance of the classifiers on the Healthy class decreases significantly, with lower recall and F1 values. The precision of the Non-healthy class also decreases, highlighting the bias of the models.

%inter-patient distribution variability
The fact that the obtained results are much lower when the data are divided in a patient wise fashion suggests that inter-patient distribution variability for the extracted metrics is very high. This stresses the need for further studies to validate the normality and non-normality values. Notwithstanding, it was already expected for the results to be lower in Scenario B, since it presents a more realistic evaluation scenario and where there is no data leakage during the training phase of the classifiers, unlike scenario A. These results underline the importance of careful experimental design, as it can have a high impact on the results that are obtained.

To understand which features were the most relevant for the classification in each scenario, the importance of each feature for the best classification model (Random Forrest) was estimated using MATLAB function \textit{predictorImportance}
\footnote{\url{https://www.mathworks.com/help/stats/compactclassificationensemble.predictorimportance.html}}. The function computes estimates of the predictors' importance for the classification model by summing changes in the risk due to splits on every predictor and dividing the sum by the number of branch nodes. \autoref{tab:features_importance} lists the 10 most important features for the best classification model in both scenarios. The importance value for each feature was determined as the average value across all 50 runs. 

\begin{table}[ht]
\centering
\caption{Ten most important features for Random Forest Model in both evaluation scenarios.}
\label{tab:features_importance}
\begin{tabular}{|c|c|}
\hline
\textbf{Scenario A} & \textbf{Scenario B} \\ \hline
ratio\_right\_left & corr\_verticalR\_verticalL \\ \hline
corr\_global\_posterior & corr\_horizontalA\_horizontalMA \\ \hline
corr\_horizontalA\_horizontalMA & corr\_horizontalMP\_horizontalP \\ \hline
corr\_verticalR\_verticalL & ratio\_right\_left \\ \hline
corr\_quadrant1\_quadrant3 & corr\_global\_horizontalP \\ \hline
corr\_horizontalMP\_horizontalP & ratio\_quadrant1\_quadrant4 \\ \hline
GI\_horizontalP & corr\_global\_quadrant2 \\ \hline
corr\_global\_horizontalP & corr\_global\_posterior \\ \hline
ratio\_quadrant1\_quadrant4 & corr\_global\_verticalL \\ \hline
corr\_global\_horizontalMA & GI\_horizontalP \\ \hline
\end{tabular}
\end{table}

\begin{figure}[ht!]
	\centering
	\includegraphics[width=0.95\columnwidth]{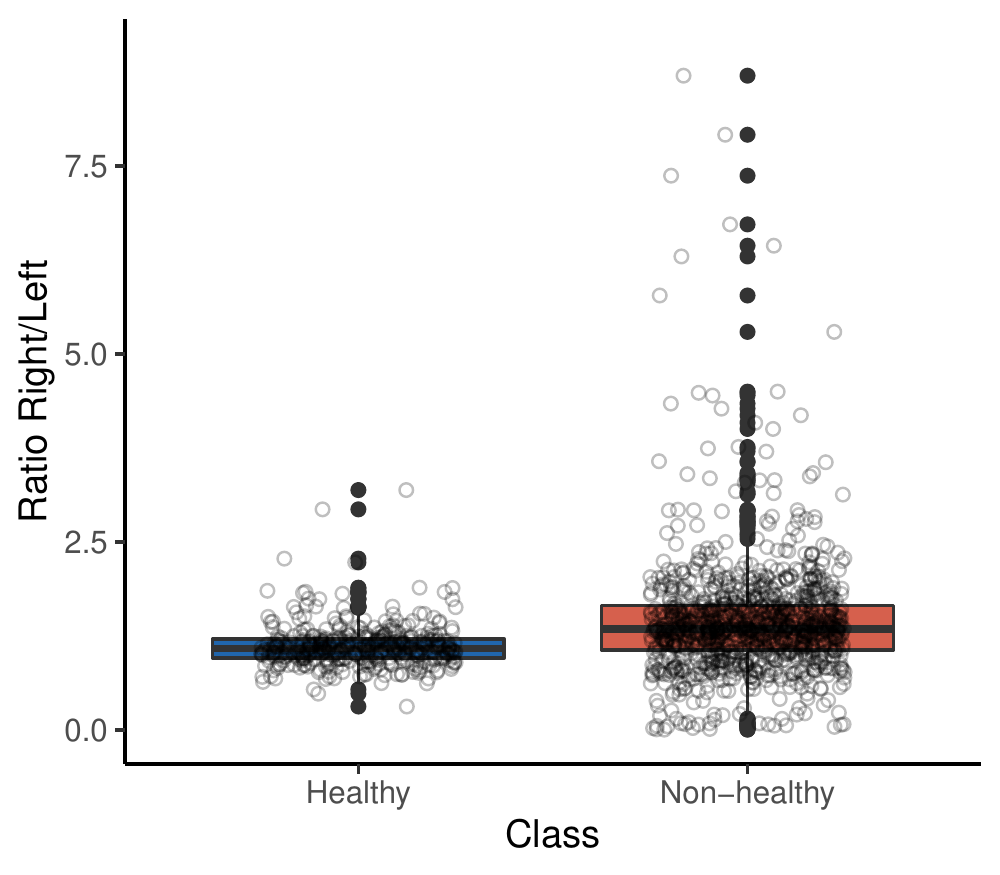}
	\caption{Boxplot with the distribution of the most important feature for the Random Forest model in scenario A (Ratio between right and left ROIs).}
	\label{fig:BoxplotScenarioA}
\end{figure}

\begin{figure}[ht!]
	\centering
	\includegraphics[width=0.95\columnwidth]{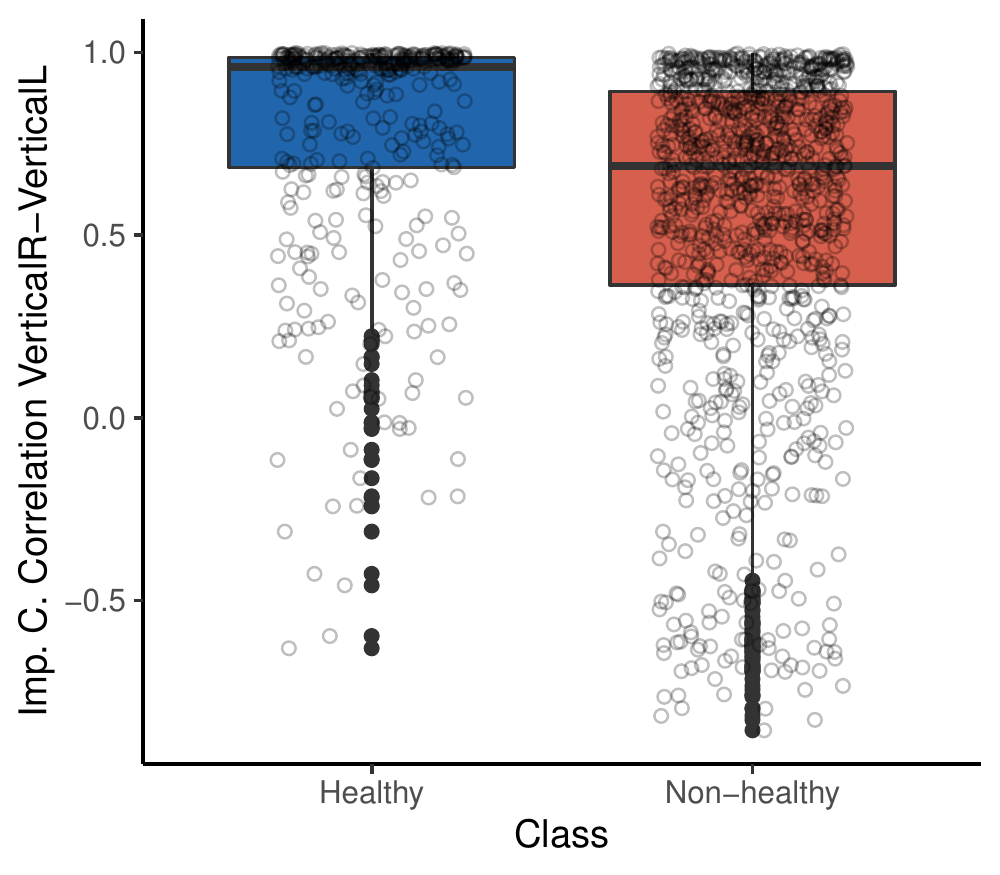}
	\caption{Boxplot with the distribution of the most important feature for the Random Forest model in scenario B (Impedance curve correlation between vertical right and left ROIs).}
	\label{fig:BoxplotScenarioB}
\end{figure}

From the analysis of \autoref{tab:features_importance}, it possible to observe that the Impedance Curve Correlation features are among the most important in both evaluation scenarios. In addition, there is a significant level of accordance regarding the most relevant features between the classification models in the two scenarios, being 8 of the 10 most important common in both cases. This fact highlights the discriminating power of those features. In \autoref{fig:BoxplotScenarioA} and \autoref{fig:BoxplotScenarioB}, the distributions for the most important features in each of the evaluation scenarios are presented (\textit{ratio\_right\_left} and \textit{corr\_verticalR\_verticalL)}, respectively). From the analysis of both figures it is noticeable that the two features present significantly different distributions between healthy and non-healthy subjects.

%\FloatBarrier

One of the major limitations of this study is related to the limited amount of data that was used and the lack of class balance. Furthermore, the characteristics of the healthy group are significantly different from the non-healthy group, with the participants in the healthy group being much younger. All of these factors combined can introduce significant bias in the classification models. Since EIT is still not a fully established imaging modality in the clinical practice, the number of studies conducted using large amounts of data using this data-source is still scarce. Moreover, the publicly available EIT datasets typically contain a small number of samples and are mostly conceived for demonstration/education purposes. 

Another possible limitation could be related to the labeling process of the samples used to develop the classifiers. If a sample was from a subject who was healthy, it was tagged as healthy. The same principle was applied to non-healthy subjects. However, even if a subject is non-healthy, this does not necessarily imply that some of the samples cannot be "normal". Even if this fact might introduce bias on the training process of the machine learning algorithms, it is very hard to avoid it in this particular case, since it would require an extensive one-by-one validation of all the samples available. Moreover, the default values of "normality" for the extracted features are not established, making this process unfeasible.

\section{Conclusion}
With this work we were able to demonstrate the potential of pairing EIT data processing and feature engineering with machine learning classification models. The obtained results, while far from perfect in more challenging evaluation scenarios, suggest that these methods might be useful to develop differential diagnosis systems using EIT and exploit the main potentialities of this imaging technique. We have also introduced a new feature/measure in EIT data processing (Impedance Curve Correlation), which has proved to be relevant to discriminate between the lung function of healthy and non-healthy subjects.

The implementation of more refined and mature methods, similar to those developed in this work, could potentially be integrated into telemonitoring applications acting as a quick diagnosis-support tool. This highlights the need to keep working on better methodologies, namely feature engineering methods that extract clinically meaningful information from the EIT data. Future work could also focus on multi-modal approaches, such as combining EIT with lung sound data. 

%Also, other machine learning approaches could be studied in future work such as multi-instance learning.
\section*{ACKNOWLEDGMENT}
This research is partially supported by \textit{Fundação para a Ciência e a Tecnologia} (FCT) Ph.D. scholarships (2020.04927.BD and SFRH/BD/135686/2018), by the Horizon 2020 Framework Programme of the European Union project WELMO (under grant agreement number 825572), and by FCT project Lung@ICU (under grant reference DSAIPA/AI/0113/2020).

\bibliographystyle{ieeetr}%ieeetr,abbrv
\bibliography{References}
\end{document}